\documentclass[12pt]{iopart}
\usepackage{epsf,graphicx,times,courier,amssymb,harvard}

\begin{document}

\title{The statistical mechanics of complex signaling networks : nerve
growth factor signaling}

\author{K S Brown\dag\footnote[5]{Present address: Molecular and Cellular Biology, Harvard University, 16 Divinity Avenue,
Cambridge, MA 02138, USA}, C C Hill\dag\footnote[6]{Present address: Gene Network Sciences, 2359 N. Triphammer Road, Ithaca, NY 14850, USA}, G A Calero\ddag, K
H Lee\S, J P Sethna\dag and R A Cerione\ddag$\|$}
\address{\dag\ LASSP, Department of Physics, Cornell University, Clark Hall, Ithaca, NY 14853, USA}
\address{\ddag\ Department of Chemistry and Chemical Biology, Cornell University, Baker Laboratory, Ithaca, NY 14853, USA}
\address{\S\ School of Chemical and Biomolecular Engineering, Cornell University, Olin Hall, Ithaca, NY 14853, USA}
\address{$\|$\ Department of Molecular Medicine, Cornell University, Veterinary Medical Center, Ithaca, NY 14853, USA}

\ead{kevin@mcb.harvard.edu}

\begin{abstract}

It is becoming increasingly appreciated that the signal transduction systems used by eukaryotic cells to achieve a variety of essential responses represent
highly complex networks rather than simple linear pathways.  The inherent complexity of these signaling networks and their importance to a wide range of
cellular functions necessitates the development of modeling methods that can be applied toward making predictions and highlighting the appropriate experiments
to test our understanding of how these systems are designed and function.  While significant effort is being made to experimentally measure the rate constants
for individual steps in these signaling networks, many of the parameters required to describe the behavior of these systems remain unknown, or at best,
represent estimates made from {\it in vitro} measurements which may not correspond to their value {\it in vivo}.  With these goals and caveats in mind, we use
methods of statistical mechanics to extract useful predictions for complex cellular signaling networks.  To establish the usefulness of our approach, we have
applied our methods towards modeling the nerve growth factor (NGF)-induced differentiation of neuronal cells. In particular, we study the actions of NGF and
mitogenic epidermal growth factor (EGF) in rat pheochromocytoma (PC12) cells. Through a network of intermediate signaling proteins, each of these growth
factors stimulate extracellular regulated kinase (Erk) phosphorylation with distinct dynamical profiles.  Using our modeling approach, we are able to extract
predictions that are highly specific and accurate, thereby enabling us to predict the influence of specific signaling modules in determining the integrated
cellular response to the two growth factors. We show that extracting biologically relevant predictions from complex signaling models appears to be possible
even in the absence of measurements of all the individual rate constants.  Our methods also raise some interesting insights into the design and possible
evolution of cellular systems, highlighting an inherent property of these systems wherein particular ''soft'' combinations of parameters can be varied over
wide ranges without impacting the final output and demonstrating that a few ''stiff'' parameter combinations center around the paramount regulatory steps of
the network.  We refer to this property --- which is distinct from robustness --- as ''sloppiness.''

\end{abstract}

\maketitle

\section{Introduction}

Recent experimental data points to the complexity underlying eukaryotic signal transduction pathways.  Pathways once thought to be linear are now known to be
highly branched, and modules formerly thought to operate independently participate in a substantial degree of crosstalk \cite{schlessingerRev00,hunterRev00}.
It is this complexity that has motivated attempts to use quantitative mathematical models to better understand the behavior of cellular signaling networks.
Models using nonlinear differential equations have often been used in the attempt to understand biological regulation in eukaryotes and have recently been
applied to the cell cycle in yeast \cite{tyson00}, circadian rhythms \cite{goldbeter95}, establishment of segment polarity in flies \cite{odell00}, and
transport dynamics of the small GTPase Ran \cite{macara02}.  Currently, a consortium called the Alliance for Cellular Signaling is embarking on a vastly more
ambitious project, coupling experiment and computation, to understand signaling in a model macrophage cell line \cite{acs02}.

Modeling of complex signaling modules presents three key challenges. (1)~Any model based on the equations of chemical kinetics will likely contain a large
number of parameters whose values will not have been determined experimentally \cite{bailey01}. Even in cases where a specific measurement has been made {\it
in vitro} (i.~e.~a binding constant or turnover rate), the value obtained may differ substantially from what would have been measured in cells if such a
measurement were possible \cite{minton01}.  Recent success has been achieved in facilitating the extraction of transcriptional rates from experimental
measurements of promoter activity \cite{ronen02}, but the problem remains extremely difficult for complex cellular signaling systems comprised of large numbers
of protein--protein interactions. (2)~Kinetic models tend to be incomplete because they often ignore many protein interactions in the hopes of capturing a few
essential ones. Thus, one ends up dealing with a ``renormalized'' model in which existing parameters absorb the effects of all the neglected parameters in the
``true'' model \cite{luss74}.  (3)~Additional difficulties arise from the fact that novel interacting proteins and new interactions for well--known players
continue to be discovered \cite{der98}, so one must have a flexible modeling methodology that can easily incorporate new information and data.  These three
challenges make the modeling of complex cellular signaling networks inherently difficult.

We believe that modeling these important biological systems cannot wait until all the rates are reliably measured, or even until all the various players and
interactions are discovered. Indeed, the most important role of modeling is to identify missing pieces of the puzzle. It is as useful to falsify models --
identifying which features of the observed behavior cannot be explained by the experimentalists current interaction network -- as it is to successfully
reproduce known results.  We choose the straightforward (but not standard) approach of directly simulating the differential equations given by the reactions in
our network. There are a variety of other approaches to modeling these networks. Systems models \cite{tysonrev} typically reduce the large number of equations
via a series of biologically--motivated approximations to a few key reactions: implementing these approximations demands deep insight that is often not
available for complex cellular signaling systems.  Boolean models are faster to simulate, but tend to be farther removed from the biology, and can be
misleading even in some simple cases \cite{guet02}. We also have chosen not to incorporate stochastic effects due to number fluctuations \cite{arkin98}, nor
are we exploring the validity of the experimentally presumed Michaelis-Menten reactions. This is not because we believe that stochastic effects are negligible
nor that Michaelis--Menten assumptions are free of limitations, but rather that we are testing the experimentalist's assumptions that stochastic effects are
not central and that the traditional saturable reaction forms are likely to be close to the real behavior in our particular system.

Our choice of simulating the full set of rate equations, and our desire to extract falsifiable predictions without prior knowledge of a large number of rate
constants and other parameters, demands that we develop new modeling tools \cite{bailey01}.  In this study we apply ideas from statistical mechanics
\cite{brown03,metropolis53,newman,hastings70,batt02} to extract predictions from a model for a regulatory signaling network important for the differentiation
of a neuronal cell line.  We show that this approach can make useful biological predictions even in the face of indeterminacy of parameters and of network
topology.  An underlying feature of the approach involves the use of Monte Carlo methods in Bayesian sampling of model spaces \cite{casellabk,hastings70}, and
such a sampling method was recently applied to a small transcriptional network \cite{batt02}. Our implementation and notation is described in the
``Experimental Section'' below, together with a brief discussion of its advantages and numerical details.  We have chosen a relatively well--studied system
--- MAPK activation in PC12 cells --- as a test case for our analysis in order to demonstrate that our methods will have broad applicability for cellular
signaling problems.  We show that our approach can make useful predictions even in the face of underdetermined parameters and uncertainty with regard to
network topology.  However, perhaps most important, our methods highlight some interesting and previously unappreciated features that we believe are
fundamentally inherent to complex cellular signaling systems.

\newpage

\section{Experimental Section}

\subsection{Data fitting}

A cost function such as that in \eref{costfunc} allows us to use automated parameter determination with the algorithm of our choice. Multiple metastable states
are the norm rather than the exception in high--dimensional nonlinear systems such as those found in models of eukaryotic signal transduction.  We use a simple
procedure to discover different minima. Fixed--temperature annealing \cite{vecchi83} is used to cross barriers between local minima, and selected parameter
sets obtained via this process are quenched using a local method such as Levenberg--Marquardt \cite{marq} or Conjugate Gradient \cite{CG}.  For the PC12 cell
network, we found a few shallow minima separated by large flat regions.  The minima showed qualitatively similar behavior but differed slightly in the quality
of fit.  All inferences from best fit parameters were performed using the best minimum found.

We note here that while cost minimization, Hessian matrix computation, and our Monte Carlo (see below) all require many cost evaluations, the methods we employ
are easily parallelizable, leading to great gains in computational efficiency.  In addition, two of us show elsewhere \cite{brown03} that if one is concerned
primarily with the identity of the stiffest few eigenvectors, an approximate hessian whose computational requirements scale only linearly (rather than
quadratically) with the number of parameters is equally suitable.

\subsection{Robustness calculations}

The calculations whose results are displayed in Figure \eref{fig:robust} were obtained as follows.  Assuming a quadratic cost space, we can write the expected
change in cost given a change in the log of one parameter
\begin{equation} C(\Delta \log p_i) = C_0 + \frac{1}{2} H_{ii} (\Delta
\log p_i)^2, \label{eqn:robust}
\end{equation}
where $H_{ii}$ is a diagonal element of the Hessian defined in \eref{eqn:hessian} and $C_0$ is the value of the cost at the best fit parameter values.  We
therefore define $H_{ii}/2$ as the inverse of the robustness of the model to a change in the i$^{\mathrm{th}}$ parameter.  Implicit in this definition is the
set of network functions/outputs in which one is interested, since these form the constraining data set and exert their effect in $H_{ii}$.  For example, the
robustness values calculated in figure \ref{fig:robust}(a) used our full experimental data set, containing 68 data points (time series for multiple protein
activities in response to both growth factors), and those of figure \ref{fig:robust}(b) used only 10 data points (the time series of Erk1/2 activation in
response to both EGF and NGF). Also necessary is a scale on which to calibrate the robustness result; we deem a parameter robust if moving it by a factor of
two causes the model probability (given by $\exp(-C)$) to decrease less than $1/e$.  In figure \ref{fig:robust}, this scale is indicated by a horizontal dashed
line.

\subsection{Ensembles}

We associate the cost in \eref{costfunc} with the energy of a statistical mechanical system.  The temperature $T=1$ is set by comparing the form of the
Boltzmann distribution to the probability of generating the data given the model if the errors in the data are normally distributed. Additionally, we use
information about the shape of the cost basins near the minima via the second derivative matrix of the cost (the Hessian) to generate the moves in parameter
space.  At finite temperature, the $B_k$'s give an entropic contribution to the cost which can be determined analytically, and it is this free energy -- cost
plus entropy from the $B_k$'s -- that we use in all thermal contexts \cite{brown03}.  From the ensemble, a mean $\langle [c(t)] \rangle$ and standard deviation
$\sigma(t)$ as a function of time are generated for each chemical concentration, given for the i$^{\mathrm{th}}$ chemical species by \begin{eqnarray} \langle
[c_i(t)]
\rangle & = & \frac{1}{N_E} \sum_{j=1}^{N_E} [c_i({\mathbf p}_j,t)] \\
\sigma_i(t) & = &  \left(\langle [c_i({\mathbf p}_j,t)]^2 \rangle - \langle [c_i({\mathbf p}_j,t)] \rangle^2\right)^{\frac{1}{2}},
\end{eqnarray}
where $N_E$ is the number of samples in the ensemble.

We start all Monte Carlo runs at the best fit parameters, though it is not absolutely necessary to do so.  We selected 704 independent parameter sets from over
15,000 sets initially generated by the Monte Carlo.  We chose independent states by first calculating the correlation time $\tau$ which we can obtain from the
lagged cost--cost correlation ($A(n) = \langle C(\mathbf p_i) C(\mathbf p_{i+n})\rangle$) function from a given Monte Carlo run. $\tau$ was defined as the
number of Monte Carlo steps required for the autocorrelator to drop to $1/e$ of its initial value. The initial $\tau$ steps of a Monte Carlo run were thrown
out, and subsequent parameter sets separated by $\tau$ samples were kept for analysis.  The ground state Hessian matrix $\mathbf{H}$ has elements given by
\begin{equation} H_{ij} = \left.\frac{\partial^2 C(\mathbf p)}{\partial \log p_i \partial \log p_j}\right|_{{\mathbf p} = {\mathbf p}^*}. \label{eqn:hessian}
\end{equation}
We diagonalize this matrix to obtain $\mathbf{V}$, the matrix of eigenvectors of $\mathbf{H}$. The eigenparameters $\alpha_j$, which are simply the coordinates
of the ensemble bare parameter sets along the eigendirections of the ground state Hessian, are given by
\begin{equation} \alpha_j = \sum_{i=1}^{N_p} V_{ji}
\log\left(p_i\right/p_i^*), \label{eigenpar}
\end{equation}
where $V_{ji}$ are the elements of $\mathbf{V}$, $p_i^*$ is the ground state (best fit) parameter value, and $N_p$ is the number of parameters (48 for the PC12
cell model).  Explicitly, according to our formula in \eref{eigenpar}, the eigenparameters are \emph{linear combinations of the natural logarithms of shifts in
rate constants}, or equivalently, ratios of rate constants raised to powers, which we can show by rewriting \eref{eigenpar} as \begin{equation} e^{\alpha_j} =
\prod_{i=1}^{N_p} p_i^{V_{ji}}. \label{eigenpar2}
\end{equation}
In either eigenparameter representation (equation \eref{eigenpar} or equation \eref{eigenpar2}), changing the combination of bare rate constants described by
$\alpha_j$ has a cost proportional to the j$^{\mathrm{th}}$ eigenvalue of the Hessian matrix, at least in the harmonic approximation.

\subsection{Mapping eigenvectors to data points}

To determine which data points are most perturbed by motion in an eigendirection, we Taylor expand the deviation of the model from the data \begin{equation}
r_i = y\left(x_i,\mathbf p\right) - Y_i
\end{equation}
about the minimum, which involves the Jacobian matrix of the deviations with respect to the model's (log) parameters.  The Jacobian's elements are given by
\begin{equation} J_{ij} = \frac{\partial r_i}{\partial \log p_j}.
\end{equation}
The product of this Jacobian and the matrix of eigenvectors
\begin{equation} d_{ik} = J_{ij} V_{kj}
\end{equation}
tells us exactly where the model/data agreement becomes poor as we move in any eigendirection. If $d_{ik}$ is large, then agreement with data point $i$ will be
changed by movement in direction $\mathbf v_k$. If $d_{ik}$ is near zero, the ability of the model to fit data point $i$ will show no sensitivity to motions in
parameter direction $\mathbf v_k$.  As expected, if index $k$ corresponds to a very soft mode, $d_{ik}$ tends to be small for all $i$.

\subsection{Cell culture and protein detection}

Rat pheochromocytoma (PC12) cells were maintained in RPMI 1640 (Cellgro, Herndon, VA) supplemented with 10\% horse serum, 5\% calf serum (both from GibcoBRL,
Gaithersburg, MD) and antibiotics/antimycotics at 1:1000. Sixteen hours prior to treatment with either EGF or  NGF (both Gibco), cells were resuspended in
serum--free RPMI. If LY294002 (Calbiochem, La Jolla, CA) was used, it was added to the medium 2 hours prior to growth factor treatment. Cells were lysed and
samples analyzed by SDS--PAGE. Proteins were transferred to nitrocellulose membranes (NEN, Boston, MA) and probed with anti--active Erk1/2 and anti--Erk1/2
(both antibodies from Cell Signaling Technologies, Beverly, MA). Detection was via chemiluminescence (ECL reagent, Amersham Life Sciences, Buckinghamshire,
England).

\newpage

\section{Results and Discussion}

\subsection{The system and model}

We have chosen nerve growth factor-- (NGF) versus epidermal growth factor-- (EGF) stimulated signaling activities in rat pheochromocytoma (PC12) cells as an
initial experimental system to test our modeling approach.  Pheochromocytoma cells have proven to be an invaluable model system in neuroscience
\cite{PC12clone}, because they express both EGF receptors (EGFR) and NGF receptors (NGFR) (specifically the high--affinity TrkA receptor) and will proliferate
in response to EGF treatment and differentiate into sympathetic neurons in response to prolonged treatment with NGF\@. It was previously reported that the
activation state of Extracellular Regulated Kinases (Erks) 1 and 2 is correlated with the cellular growth state of PC12 cells \cite{traverse92}. A transient
activation of Erk1/2 has been associated with EGF treatment and cell proliferation, while a sustained activity has been linked to NGF stimulation and
differentiation.  Sustained Erk1/2 phosphorylation has been suggested to be sufficient for PC12 cell differentiation \cite{robinson98}. It has since been
recognized that while both EGF and NGF receptors activate the GTP--binding protein Ras, the distinct cellular outcomes triggered by these growth factors must
lie in the differential activation of other pathways that modulate Erk1/2 activity, and several hypotheses have been proposed to account for this signaling
specificity \cite{kao01,york98,yasui01,rapp96,brightman00}. Figure \ref{fig:model} shows the topology of a model for this process (additional details are
provided in the supplemental material). The model includes a common pathway to Erk through Ras shared by both the EGFR and NGFR, but also includes two side
branches that we hypothesized were important in modulating signaling. One, through phosphatidylinositol 3-OH kinase (PI3K), can serve to downregulate Erk via
the negative regulation of Raf1 \cite{zimmermann99,rommel99}, and a second, putatively through the small GTP--binding protein Rap1 and the kinase B-Raf, can
upregulate Erk by boosting Mek activation \cite{ohtsuka96,york98,bos01}. The model presented in figure \ref{fig:model} is described by a set of 28 nonlinear
differential equations (supplemental material) and has 48 rate parameters\footnote[1]{We include the saturation $K_m$s with the rate constants.} for which
precise measurements are largely unavailable.

The growth factor-stimulated dimerization of EGFRs and NGFRs is not explicity depicted in the model, but because it is directly linked to the triggering of the
downstream signaling pathways, it is an implicit component of the receptor activation steps in our analysis.  The same is true for the adaptors (e.g. Grb2 for
growth factor receptor-binding protein 2) that interface the EGFRs and NGFRs with mSos (for mammalian Son-of-sevenless).  While there are several points in the
model where the signaling activities are subject to negative regulation, for example through the actions of GTPase-activating proteins for Ras and Rap1 and
phosphatases for Raf1, B-Raf, MEK1/2 and ERK1/2, we recognize that there are other points in the network where negative regulation can also occur (e.g. at the
level of Akt/PKB or p90/RSK).  These are not specifically depicted in the model shown in Figure 1, because we assumed that these additional negative regulatory
steps would not influence our interpretation of the time series of EGF- and NGF-stimulated ERK1/2 activation.  The same is true for EGFR down-regulation which
occurs on a slower time scale than the time course for the EGF-dependent stimulation of the Ras-Raf-MEK-ERK pathway.  However, it is relatively easy to
incorporate these or other steps that might be identified in the future as being potentially important.  Moreover, as alluded to above, we firmly believe that
the most important goals of our analyses are to establish if useful predictions for complex signaling networks can be extracted using the methodology described
here, as well as determine just how well a particular network explains available experimental data, or if in fact additional interactions and signaling
participants are necessary.

\subsection{The ensemble approach to modeling complex signaling networks}

In the analysis presented here, we primarily use the ensemble method to match the model to time courses of the activities of signaling molecules, although the
method can easily incorporate agreement with parameter measurements as well.  Time series have two major benefits: one, they are often more plentiful than
measurements of kinetic parameters, and two, they are independent of the mathematical form of the model (for example, particular kinetic schemes).  Because of
the significant degree of indeterminacy in the model, a single set of parameters forms an incomplete description.  Thus, a thermal Monte Carlo
\cite{metropolis53,newman,hastings70,batt02,brown03} is used to generate an {\it ensemble} of parameters weighted by cost, from which we can compute a mean and
standard deviation for the activity of each protein (see ''Experimental Section'' as well as the discussion of figure \ref{fig:ensemble}(a), below).  Of
particular importance is the standard deviation of the activity, because it is a direct measure of how perturbations of the parameters affect predictions of
the model. The ensemble method simultaneously makes the model more useful and falsifiable. If the variation in a particular chemical activity is large across
the ensemble, then predictions based on that outcome are unreliable -- the model is not sufficiently constrained to make a useful statement about such a
situation. Conversely, when a single parameter set is used to characterize the model, it might accommodate new data simply by wiggling some of the parameter
combinations, and thereby provide an unreliable description of the system. When using the ensemble approach, new data that falls far outside the ensemble
prediction illustrates a feature of the system that the model is incapable of representing \footnote{These methods identify typical members of the ensemble,
however, not all members of the ensemble. New data outside the original error bars may in principle be consistent with the model but restrict the parameters to
a previously insignificant subregion.}.

Figure \ref{fig:ensemble}(a) shows the results of the ensemble approach applied to the PC12 cell system.  The figure illustrates the active Erk1/2 response to
EGF and NGF\@.  In all, fourteen chemical times series from seven experiments performed in four laboratories were used for ensemble generation (see
supplemental material).  As discussed above, rather than displaying a single fit, an average over many fits is presented, for which the details are as follows.
To quantitatively compare the model's output to data, a least squares cost function is used
\begin{equation} C(\mathbf p) = \frac{1}{2} \sum_{i=1}^{N_R} \left(
\frac{B_k y\left(x_i,\mathbf p\right) - Y_i}{\sigma_i}\right)^2, \label{costfunc}
\end{equation}
where $Y_i,\sigma_i$ are the value and error of the i$^{\mathrm{th}}$ data point, $y(x_i,\mathbf p)$ is the corresponding model output evaluated with parameter
set ${\mathbf p}$, and $N_R$ is the total number of experimental data points.  In particular, a typical $p_i$ might be a binding constant for a
receptor--ligand interaction or a Michaelis constant for an enzymatic step, while a data point $Y_i$ might be the concentration of a phosphorylated protein at
$x_i = 30$ min (see figure \ref{fig:ensemble}(a)) with error bar $\sigma_i$, and $y(x_i,\mathbf p)$ would be the theory curve for that same phosphorylated
protein at the same time. We introduce the factor $B_k$ for the $k^{\mathrm{th}}$ chemical species\footnote{While the word ``chemicals'' could denote proteins,
RNA, small molecules, etc., for purposes of this study all the chemicals are proteins, sometimes separated into different phosphorylation or GTP--binding
states.}, which is an overall scaling factor for the theory curve, in order to accommodate experimental data sets with a variety of units.  It is worth
emphasizing that while time series measurements are the focus of this study, virtually any type of data (for example, dose--response information or
experimental measurements of parameter values) can easily be incorporated into such a framework.

The theory curves in figure \ref{fig:ensemble} are given by the continuous solid and dotted curves; in this and subsequent figures showing ensemble activation
curves, the central curve is the ensemble mean and the curves surrounding it show one standard deviation. It is clear from figure \ref{fig:ensemble}(a) that
the model reproduces the expected activation of Erk1/2 by EGF and NGF\@.

\subsection{Implications for modeling complex signaling networks: Inherent sloppiness of signal transduction}

Our analysis, within the context of the PC12 cell system, yields a number of interesting implications regarding the complex networks used to propagate cellular
signals.  One such implication becomes evident when considering to what degree particular {\it combinations} of rate constants (rather than single rate
constants) can be shifted in our model and still yield a good description for the Erk activity profiles.  We refer to this property as ``sloppiness.''

A simple example is as follows.  A standard first-order protein binding reaction may be characterized by two rate constants: $k_{\mathrm{on}}$ and
$k_{\mathrm{off}}$.  However, in many contexts it is not these individual rate constants that are most relevant but instead the equilibrium constant $K_d =
k_{\mathrm{on}}/k_{\mathrm{off}}$, a measure of the affinity of the interaction. If indeed only the ratio of the on and off rates matters, then their
\emph{product} will be irrelevant: both rates can be increased by the same factor (a twofold increase in both, for example) and $K_d$ will not change. For such
a case, we would call $K_d$ ``stiff,'' because it cannot be changed without noticeable biological effects, while the product $k_{\mathrm{on}} k_{\mathrm{off}}$
will be ``soft,'' since it can be changed freely.

Rather than arbitrary ratios and products of rate constants, we analyze the fluctuations in a particular set of rate constant combinations: the eigenvectors of
the Hessian matrix of the cost, expressed in terms of the logs of the rate constants (see ``Experimental Section'').  These particular vectors in parameter
space and the degrees to which they can change while still preserving the appropriate biological response give us information about key degrees of freedom in
the model. A paramount feature of the eigenvectors is that they are a unique set of alternate model parameters that show no covariance; they are independent in
a way that single rate constants and other non-eigenvector combinations are not.  Unlike in the simple example given above, the eigenvectors are typically
combinations of more than two rates (see ``Experimental Section''), but one can interpret them in a similar light: the eigenvectors are particular combinations
(ratios and products) of rate constants that may be individually varied. Changes in the stiff directions cause large perturbations in the system's behavior
while changes in the soft directions are imperceptible.

We note here that analysis of a system in terms of eigenparameters is standard practice in physics, and we feel it is especially useful when analyzing models
of signal transduction for a number of reasons.  First and perhaps most importantly, the character of the eigenparameter fluctuations (figure
\ref{fig:ensemble}(b)) is what allows us to make predictions. While individual rate parameters flop around hopelessly, we see in the stiff eigenparameters that
\emph{some} degrees of freedom are reasonably well-constrained by the data.  Our ability to determine these few stiff eigenvectors well is what allows us to
make any predictions at all, even in the face of many unknowns (also, see the ``Robustness'' section and figure \ref{fig:ellipse} for more discussion on this
point). Secondly, and as will be considered further below (see the next section) the eigenparameters can have a physical interpretation that can decompose the
system into dynamical modules (this kind of interpretation is similar to the meaning given to the dissociation constant above). Third, analysis of the
eigenparameters and their corresponding eigenvalues allows us to see similarities among seemingly unrelated models (Brown and Sethna, in preparation).

An eigenvector analysis is shown in Figure 2B which plots linear combinations of the natural logarithms of shifts in rate constants (see ``Experimental
Section'' for a full description).  The relatively large size of the vertical bars for the majority of the eigenparameters in figure~\ref{fig:ensemble}(b)
(note the natural log scale of the vertical axis) indicates that only a small fraction of parameter combinations are well--constrained by the data (these are
the aforementioned ``stiff modes'' --- also, see next section).  The majority of the eigenparameters (around 80\%) are ``soft'' and vary over more than an
order of magnitude and sometimes many orders of magnitude. The large number of soft directions exemplified by figure~\ref{fig:ensemble}(b) leads us to refer to
the modeling of complex signaling networks as \emph{sloppy models}, and we emphasize that this softness does not arise from a trivial underdetermination of
parameters, as 59 parameters (rate constants plus $B_k$'s in \eref{costfunc}) were fit to 68 data points in figure \ref{fig:ensemble}. While there are many
protein activities for which we have no direct data, it is important to point out that the degree of sloppiness is preserved even when we construct an
artificial situation by generating an abundance of ``perfect data'' from the model itself \cite{brown03}. It is in fact because of the inherent sloppiness of
cellular signaling systems that we confine our predictions to chemical activities and not values of rate constants. Figure \ref{fig:ensemble}(b) provides a
strong argument for the ensemble approach, as the huge variation in eigenparameters makes description of a complex cellular signaling network with a single set
of ``best'' parameters perilous.

\subsection{Implications for modeling complex signaling networks: Only a few parameter combinations are ``stiff''}

Another important implication for complex cellular signaling networks that emerges from our analysis is that only a small fraction of parameter
\emph{combinations} are important in determining the behavior of the signaling system that we are studying, and one may naturally ask: what are these
combinations? Which rate constants appear in the ``stiff degrees of freedom''?  The stiffest mode corresponds approximately to the parameter combination
($k_{\mathrm{GAP}} k_{\mathrm{dRaf}}/k_{\mathrm{Sos}} k_{\mathrm{Ras}} K_{\mathrm{mSos}}$), which is a ratio of rate constants that \emph{antagonize} Ras and
Raf function (e.~g.~the rate for GAP--catalyzed deactivation of Ras and the deactivation rate for Raf) to those that \emph{promote} it (activation of Ras by
Sos, activation of Raf by Ras, and the $K_m$ for RasGAP--catalyzed GTP hydrolysis). This points to Ras and Raf as the critical nexus for generating the
appropriate signaling behavior.  Perhaps not surprisingly, the proteins which emerge as the key control points in our model are indeed those, when mutated,
that are most likely to cause disease.  We emphasize that we have arrived at this conclusion --- that Ras and Raf are key in the network --- solely through the
coupling of the model to time series data for a variety of proteins, not just Ras and Raf.  This points to the power of stiff mode analysis in highlighting key
proteins and interactions, which would be particularly valuable in cases where such key regulators are \emph{not} yet known experimentally.  The second
stiffest mode predominantly involves rates that localize to the feedback loop from Erk to p90/RSK to mSos (Figure 1); it is the ratio of the $K_m$s for p90/RSK
activation and Sos deactivation to the same activation/deactivation rates ($k_2$s). This highlights negative feedback from Erk to Sos as a second key point for
regulation in the network.

We can go beyond simply identifying those parameter combinations that need to be tightly constrained to asking the question: if we move in a stiff direction,
where does the model begin to miss the data most significantly? We investigate this by Taylor expansion of the deviation of the model from the data (see
``Experimental Section''). We find that the stiffest mode --- the one localizing to Ras and Raf --- is most important in achieving the correct fast response,
on the scale of a few minutes, for many proteins under the regulation of both growth factors. The second stiffest mode when perturbed, on the other hand, has a
dramatic effect on Ras activity on longer timescales (thirty minutes) and also regulates Rap1 activity (i.~e.~the second stiffest mode mixes rates from the
Rap1 loop with those in the feedback loop from Erk to mSos). Thus, the identity of the stiff eigenvectors points to key control points in the signaling
network. Moreover, because the dominant rate constants in the stiffest eigenvectors do not have to appear close to each other in the network, the stiffest
eigenvectors can identify ``dynamical modules'' that mix multiple static protein modules. Finally and perhaps most important, mapping eigenvector perturbations
back to the data shows what aspects of the temporal profile are affected by disruption of these critical regulatory groups and can suggest manipulations that
would ``tune'' the system to a particular response.

\subsection{Implications for modeling complex signaling networks: Robustness}

Using the ensemble approach to analyze the PC12 cell signaling model also allows us to ask questions about the sensitivity of the network's activity to changes
in single (rate) parameters. In the protein network modeling literature, this property is called robustness, defined as the buffering of a biological network's
function to changes in its components, rates, or inputs \cite{leibler97} and this property is usually tested by looking for changes in one or two of a model's
protein outputs in response to changing single rate constants \cite{odell00,macara02,laubloomis98}. Figure \ref{fig:robust}(a) shows our results for the
robustness of the PC12 cell network, with details for the calculation presented in the figure legend and in ``Experimental Section.''  The dotted line in the
figure sets the value of a calibration scale; every bar falling below the dotted line corresponds to a robust parameter, and those falling above are nonrobust.
In addition, the relative robustness or nonrobustness of a parameter is given by the height of its corresponding vertical bar --- a bar that is half as high as
another indicates the former parameter is twice as robust as the latter. We can see that when we take into account the activities of all the proteins included
in our full experimental data set, the model is quite nonrobust, since changes in 80$\%$ of the parameters result in significantly worse agreement with data
(figure~\ref{fig:robust}(a)).

The fact that our model is not very robust is both because we have a quantitative measure of model agreement (equation \eref{costfunc} rather than ``fit by
eye'') and because we consider more than a single output or behavior of the model, since we have data for the active forms of six of the proteins of the PC12
cell model, some for both EGF and NGF treatment. Thus, we have a rather stringent criterion for model success.  However, this picture changes dramatically if
we consider the system in terms of a simple ``input/output'' relationship.  For example, if we consider only Erk activation by EGF and NGF as the pertinent
outputs, we see a much more robust response, finding in this case only 9 nonrobust parameters rather than 37 (see figure \ref{fig:robust}(b)).  While in some
cases viewing biochemical systems as input/output transducers is justified, in others it is not a realistic description.  For example, the Ras protein has a
myriad of cellular targets \cite{der98} and in many cases, the physiological roles of Ras in the cell may require its stimulation of multiple targets/effectors
as well as complex ''cross-talk'' between these different effectors.

Overall, we feel that our simple definition of robustness offered in \eref{eqn:robust} (''Experimental Section'') and the subsequent discussion gives a
quantitative measure of what one means by robustness at any level of detail, whether one is interested in input/output characteristics (figure
\ref{fig:robust}(b)), the behavior of several internal network components (figure \ref{fig:robust}(a)), or a level of description somewhere in between (two
outputs rather than one, for example). However, it is important to appreciate the differences between the robustness of a parameter, a parameter's error bar
(as obtained from the formal covariance matrix of the fit \cite{nr}), and the ranges of the eigenparameters (as shown in figure \ref{fig:ensemble}(b)).  When
we examine the error bars on the fitted parameters for the PC12 cell model, we find them to be huge: the most well-defined rate constant has an error bar of
10$^3$.  The schematic in figure \ref{fig:ellipse} provides an example of the differences that can occur with regard to the relative robustness versus the size
of the size of the error bars for a simple two--parameter case, where in this example the range for the error bars (schematically indicated by the red dotted
lines and arrows labeled ``s1'' and ``s2'') is much larger than the typical ``robustness range'' (illustrated by the blue axes with tips labeled ``r1'' and
``r2'').  This example, described in more detail in the legend to Figure 4, immediately begs two questions: (1) How do we understand the enormous error bars on
the rate constants? (2) How can we hope to generate {\it any} useful predictions from models of cellular signaling systems, given the large ranges on parameter
values that can and do occur?

Figure \ref{fig:ellipse} helps to provide answers to both of these questions.  First, we see that there is no contradiction in having a parameter's error bar
be much larger than its robustness range; as the ellipses in figure \ref{fig:ellipse} become more stretched, the error bars on the parameters grow while their
robustness ranges remain roughly the same.  Secondly, and more importantly, the eigenvectors illustrate how it is that we can make any predictions at all.
While the {\it individual} rate constants may all have huge error bars, the same need not be true of the eigenvectors. Some eigenvectors ($v_2$ in figure
\ref{fig:ellipse}) can be even more poorly determined than the worst-constrained single parameter, but some (like $v_1$) can be much better determined than
even the most well-constrained parameter in the model.

Thus, in summary the parameter ranges obtained from our robustness calculation assume that a parameter value is changed singly while others remain fixed.  In a
formal covariance calculation, the fact that other parameters may shift to compensate for part of the change in a single parameter leads to larger error
ranges: for example, there may be more variation because other parameters are able to offset a change in a given rate constant.  The eigenvector ranges also
assume more than one rate constant can be moved at once, but eigenvectors are very particular combinations: motions in one eigendirection \emph{cannot} be
offset or compensated by motions in any other eigendirection.  In practical terms, the questions of most pressing biological interest are not rate constant
values but changes in cellular behavior upon different interventions and mutations.  We will see in the next section that these biologically relevant
quantities are well predicted by our model, despite the fact that we have shown that we cannot predict reaction rate constants with our time series data.

\subsection{Predictions and experimental verification}

An important goal is to use our model to make testable predictions about biological or cellular experiments for which we have no results. For the PC12 cell
signaling network, we have hypothesized that the left arm of the pathway (see figure \ref{fig:model}) is suppressing Erk1/2 activation by EGF\@. However, what
do we predict when PI3K is not activated? If PI3K activation is critical to the transient activation of Erk1/2 by EGF, then inhibition of PI3K might cause
sustained Erk1/2 activation and give rise to cellular differentiation.  If this were true, then it should be possible to convert EGF to a differentiating
factor using the PI3K pathway. Figure \ref{fig:predict}(a) shows an ensemble prediction for the time series of Erk1/2 in response to EGF, under conditions
where PI3K activity has been inhibited.  We were quite surprised to find that the model predicts the opposite of our initial expectation, namely that PI3K
inhibition will not lead to differentiation in response to EGF\@. We emphasize that this is not a prediction of the single best fit parameters: this approach
examines a sampling of all parameters consistent with available time-series data.  It is particularly noteworthy that precise predictions of biologically
relevant time series data can be reliably extracted from a model whose rate constants are each varying over many orders of magnitude.

Figure \ref{fig:western} shows the results obtained from experiments using LY294002, a pharmacological inhibitor of PI3K. The qualitative agreement between
model and experiment is quite good, and both clearly show that the inhibition of PI3K activity neither produces a sustained Erk1/2 signal in response to EGF
nor does it switch the NGF--induced Erk1/2 activation profile from a sustained to a transient response (although it does provide some quantitative tuning of
the signal at short times). Also, the inhibition of PI3K activity does not cause significant morphological differentiation in combination with EGF treatment
(supplemental material), as expected from the relationship between sustained Erk1/2 activity and differentiation.

In particular, these analyses suggest that the PI3K to Akt loop in figure \ref{fig:model} (shown as grayed--out) is not necessary to account for the
differences in Erk1/2 activation in response to EGF versus NGF\@. This is further supported by the cluster of ``irrelevant'' rate constants near rate constant
number 30 in figures \ref{fig:robust}(a) and \ref{fig:robust}(b), since they are all associated with this loop. Indeed, such a reduced model, generated by
removing the PI3K to Akt loop, describes the Erk1/2 data (and the rest of the data used in this study) only slightly worse than the larger model, yielding an
optimal cost that is about $30\%$ percent higher, and similar ensemble predictions.

We also make predictions about manipulations that, unlike inhibition of PI3K, do dramatically affect Erk activation.  Figure \ref{fig:predict}(b) shows the Erk
response in PC12 cells to 50 ng/ml NGF and either dominant negative (DN) Ras (blue curve) or Rap1 (red curve), a member of the Ras family of GTP--binding
proteins that like Ras is activated by the NGF receptor.  These predictions show good qualitative agreement with previous experiments \cite{york98}, in which
DNRas interferes with early Erk activation but not its eventual sustained behavior and DNRap1 affects the long--term value of Erk phosphorylation but not its
early activation. Thus, taken together, Figures \ref{fig:predict} and \ref{fig:western} show that our model agrees with, and can predict the results of,
experimental manipulations that disrupt each of the three main pathways we have included in our model: PI3K, Rap1, and Ras.

\section{Conclusions and Outlook}

In conclusion, motivated by ideas from physics, we have used a formalism (part of which is termed elsewhere the ensemble method (Battogtokh et al., 2002)) for
modeling cellular signaling networks that (1) provides biologically accurate descriptions of the signal output, and (2) is falsifiable even in the face of a
high degree of uncertainty regarding the rates and binding affinities for many of the steps comprising complex biological systems. We have applied this
methodology to the signaling systems that underlie NGF-induced neurite extension and differentiation of PC12 cells and have been able to evaluate the
importance of different regulatory loops in generating the key signaling endpoint, the sustained activation of Erk. However, what may be most important, our
modeling efforts have yielded some interesting and perhaps previously unappreciated implications and lead us to emphasize some key points regarding complex
cellular signaling networks. First, it is clear that only a small fraction of parameter {\it combinations} (the eigenparameters) for such signaling systems are
likely to be well-constrained, and most if not all individual rate constants can vary over huge ranges.  In fact, some of the error bars for the individual
rate constants can be enormous even though the sensitivity of the system to a change in these parameters is high (i.e. even when the parameters are not robust)
due to covariance of other rate constants that can compensate.  Second, the few well-constrained (stiff) parameters reveal critical focal points in the
signaling network for ensuring the generation of an appropriate output; in the case of the PC12 cell system, the stiffest mode encompasses the regulation of
Ras and Raf, two proteins which are well known for playing crucial regulatory roles in mitogenic pathways, as well as in cellular differentiation, and when
mutated, stimulate oncogenic transformation.

Overall, this now leads to our appreciation that complex signaling systems are characterized by an inherent sloppiness, such that significantly marked
variations in different combinations of parameters can yield similar outputs and end results. This inherent sloppiness offers some intriguing possibilities
regarding the way in which signaling systems with multiple interacting pathways can evolve.  Indeed, we believe our analysis yields insights into one way in
which properties like robustness and evolvability can simultaneously coexist.  Soft modes show a property similar to that of robust rate parameters, as soft
modes can absorb large changes without altering cellular behavior.  On the contrary, even small motions in the stiff modes can elicit large effects in cellular
signaling. Hence, when considering the effects of multiple genetic mutations, some combinations (the stiff modes) allow the organism to adapt to new conditions
while others (the soft modes) leave critical signaling activities intact and unperturbed.

\newpage

\ack K. Brown would like to thank M. Antonyak and J. Zollweg for technical assistance and G. Hoffman, D. Schneider, C. Myers, B. Ganem, and J. J. Waterfall for
helpful discussions.  We would like to thank NSF DMR-0218475 and NIH T32-GM08267 for financial support and the Cornell Theory Center for computational
resources.

We thank both the NSF and NIH for support, and we are also grateful for the resources provided by the USDA and the Cornell Theory Center.

\newpage

\bibliography{refs_brown_txtfigs}

\bibliographystyle{jphysicsB}

\newpage

\begin{figure}[h]
\begin{center}
\includegraphics[width=8.7cm]{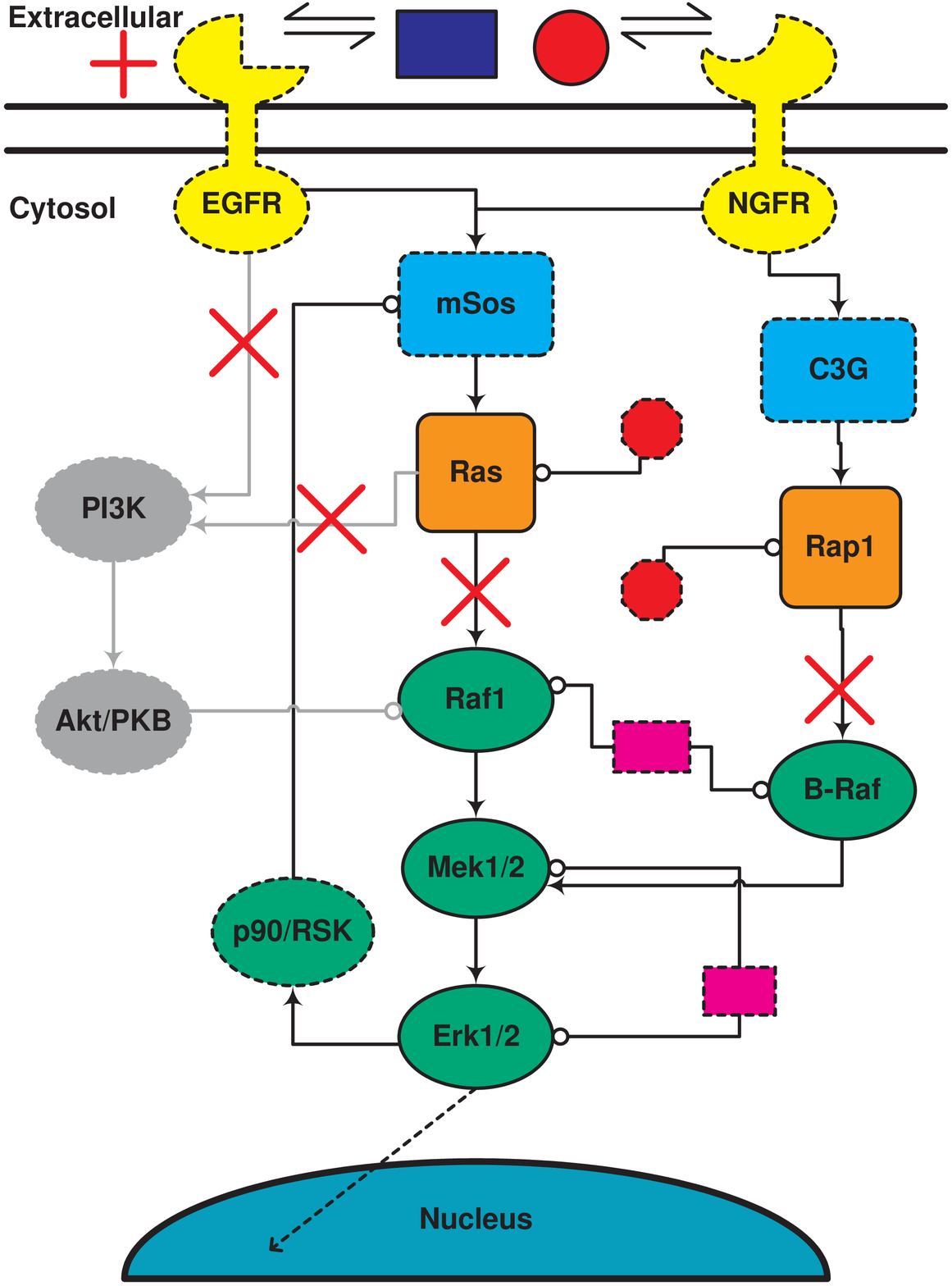}
\end{center}
\caption{Model of Erk1/2 activation by EGF and NGF in PC12 cells. EGF and NGF both activate Erk via Ras. EGF can also use the left branch involving PI3K to
modulate Erk activity through Raf1 downregulation, and NGF can upregulate Mek using the right branch containing Rap1. Double arrows indicate binding/unbinding
reactions. Single arrows indicate stimulatory effects and lines capped with open circles represent negative regulation, which in this model are all of the
Michaelis-Menten type \cite{stryer}. Small purple boxes are unregulated phosphatases and small stop signs are unregulated GTPase activating proteins (GAPs). A
solid border around a chemical indicates data for that molecule's activity was used to constrain the model, and a dotted border indicates none was available.
Red crosses indicate links that were cut to make predictions (see text).  Additional details can be found elsewhere (see supplemental material).}
\label{fig:model}
\end{figure}

\begin{figure}[h]
    \begin{center}
        \begin{tabular}{lc}
     {\bf \Large (a)} & \resizebox{8.8cm}{!}
      {\includegraphics[width=8.7cm,angle=-90]{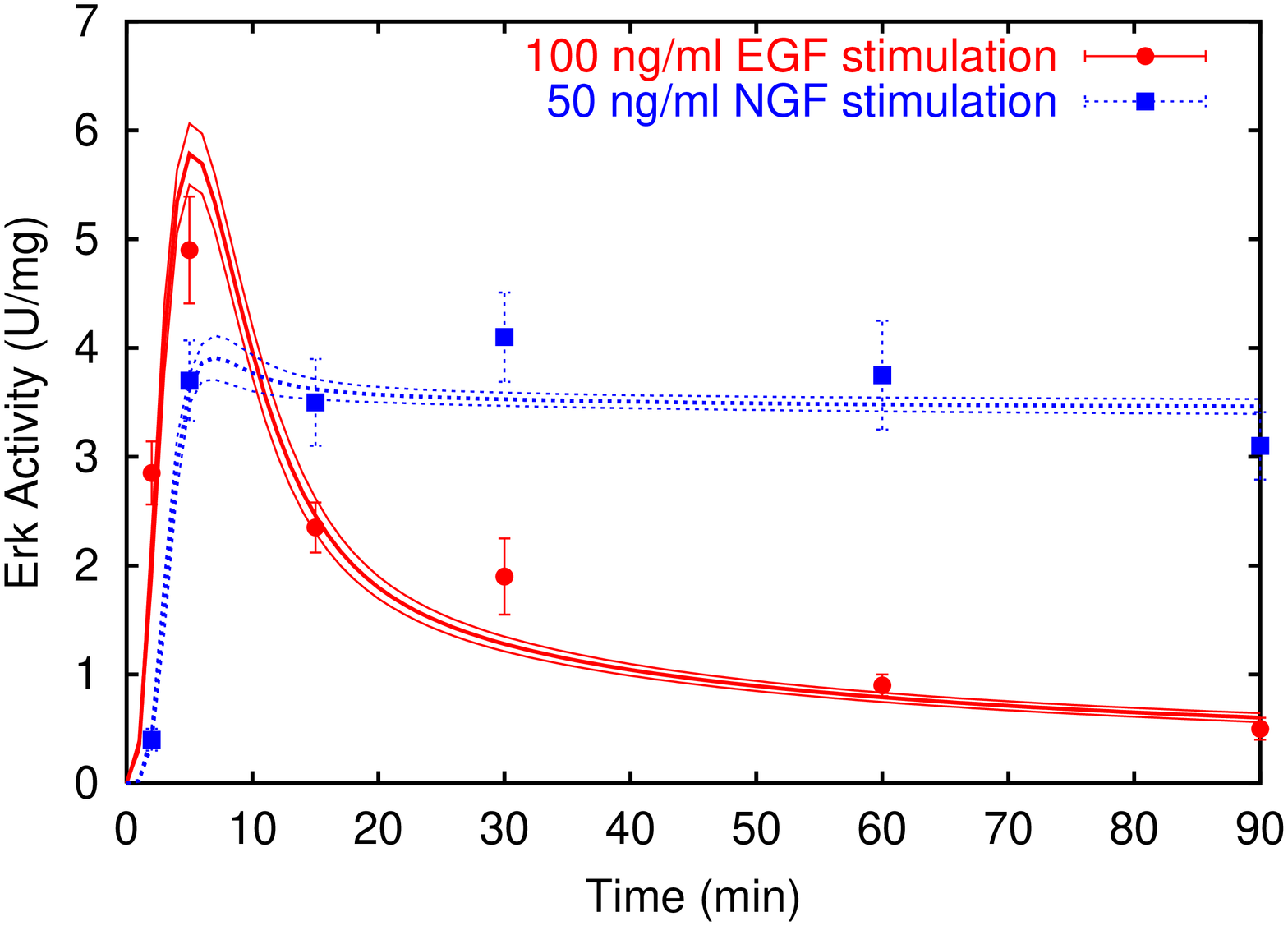}} \\
       \vspace{0.25in} \\
     {\bf \Large (b)} & \resizebox{8.8cm}{!}
      {\includegraphics[width=8.7cm,angle=-90]{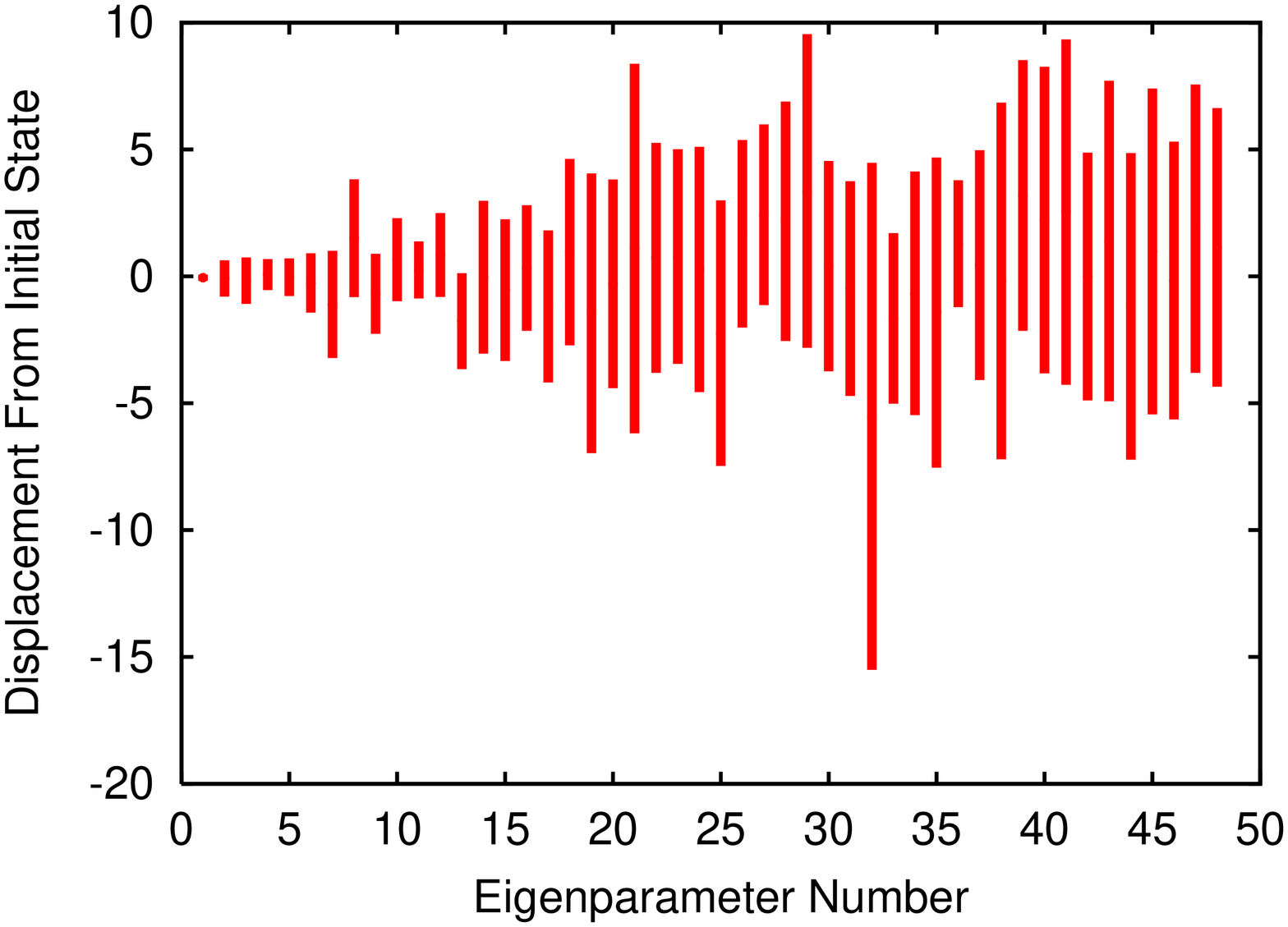}} \\
    \end{tabular}
\end{center}
\caption{Behavior of the PC12 cell model. (a)~Ensemble of acceptable solutions for the model with topology given in Fig.~1. Means and sample standard
deviations are calculated as described in the text from a 704--member ensemble of independent samples. Together with the data shown \cite{traverse94}, fourteen
data sets from seven experiments performed in four laboratories were used in ensemble generation (see supplemental material). The error bars here and in
subsequent figures represent one standard deviation, or the sixty--six percent confidence level (in contrast to the four standard deviation ranges in
\citeasnoun{batt02}). Notice that EGF stimulates a transient Erk response (leading to proliferation) and NGF stimulates a sustained response (leading to
differentiation). (b)~Scatterplot of the eigenparameters (defined in \eref{eigenpar}) for the 704 member ensemble. The eigenparameters from the ensemble
densely populate the area covered by the colored bars, which extend from the minimum value in the ensemble to the maximum value.  The scale for the $y$ axis is
the natural logarithm (base $e$).  Stiff eigenparameters exhibit small fluctuations (short bars) and soft eigenparameters large fluctuations (tall bars).
Notice that while the eigenparameters vary over several orders of magnitude, the variation in Erk1/2 activity shown in~(a) is small. The variation of the
softest eigenvalues in complete equilibrium is likely even larger.} \label{fig:ensemble}
\end{figure}

\begin{figure}[h!]
    \begin{center}
        \begin{tabular}{lccc}
      {\bf \Large (a)} & \resizebox{40mm}{!}
     {\includegraphics[]{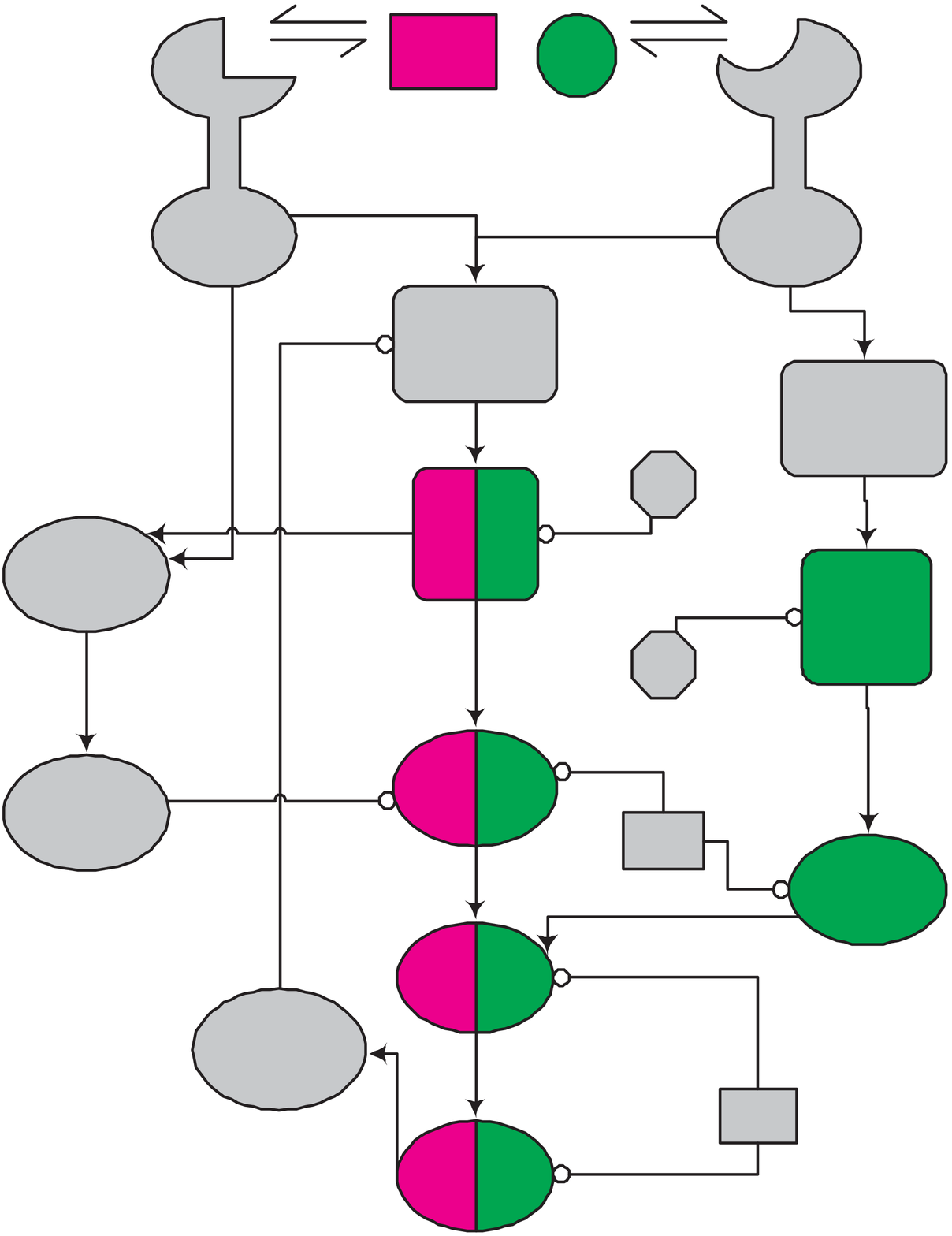}} & \hspace{0.5in} & \resizebox{88mm}{!}{\includegraphics[]{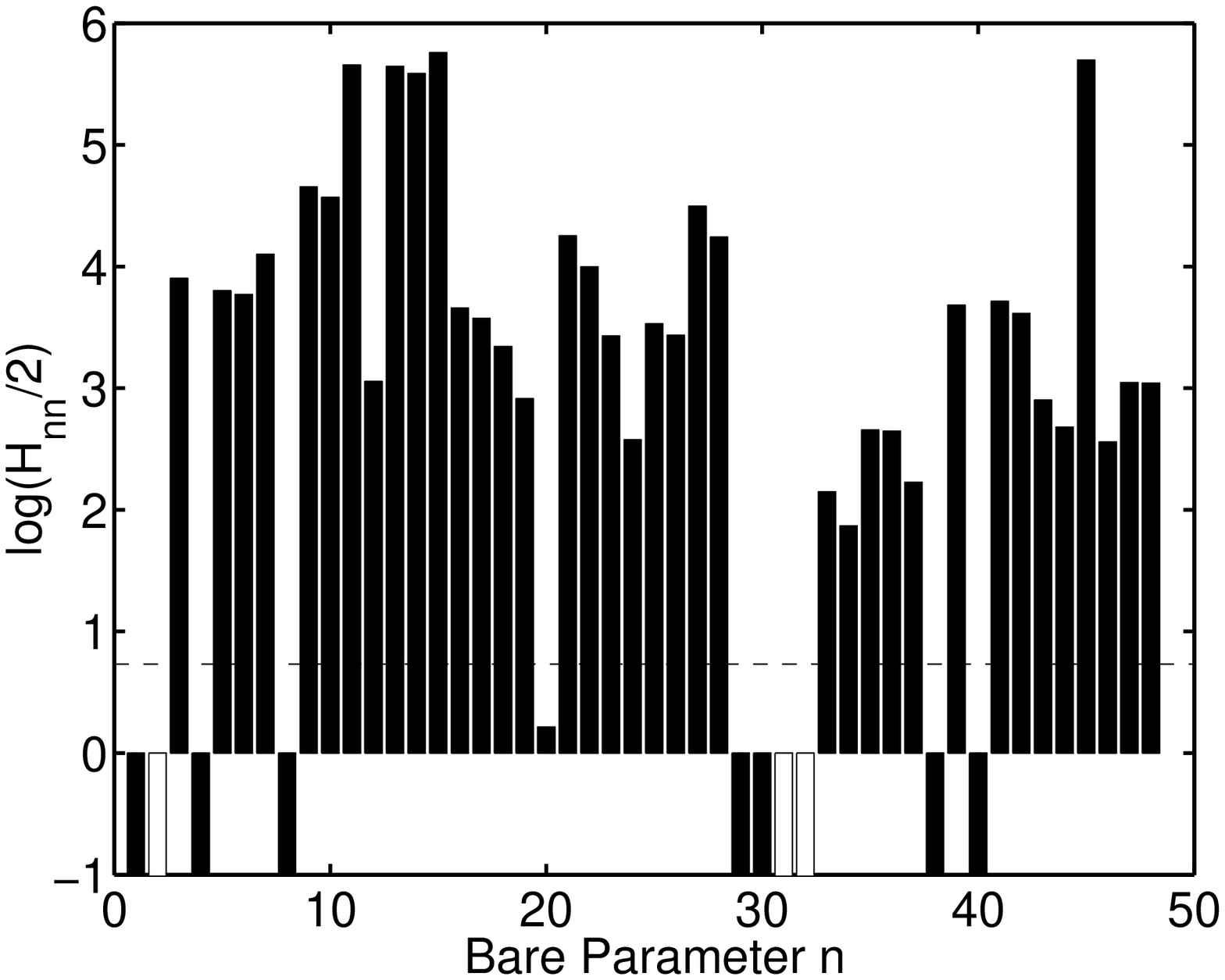}}\\
        \vspace{0.15in} \\
      {\bf \Large (b)} & \resizebox{40mm}{!}
      {\includegraphics[]{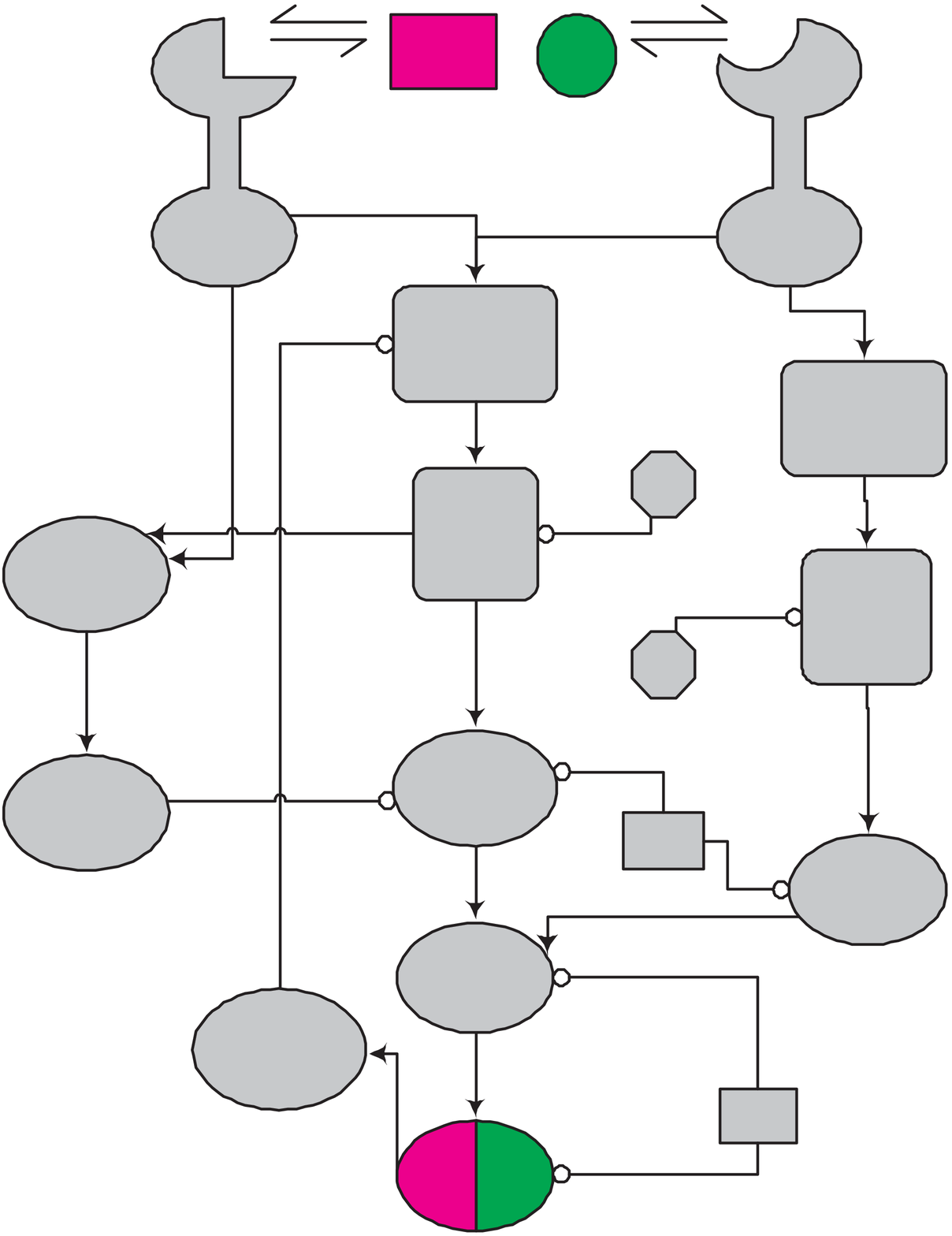}} & \hspace{0.5in} & \resizebox{88mm}{!}{\includegraphics[]{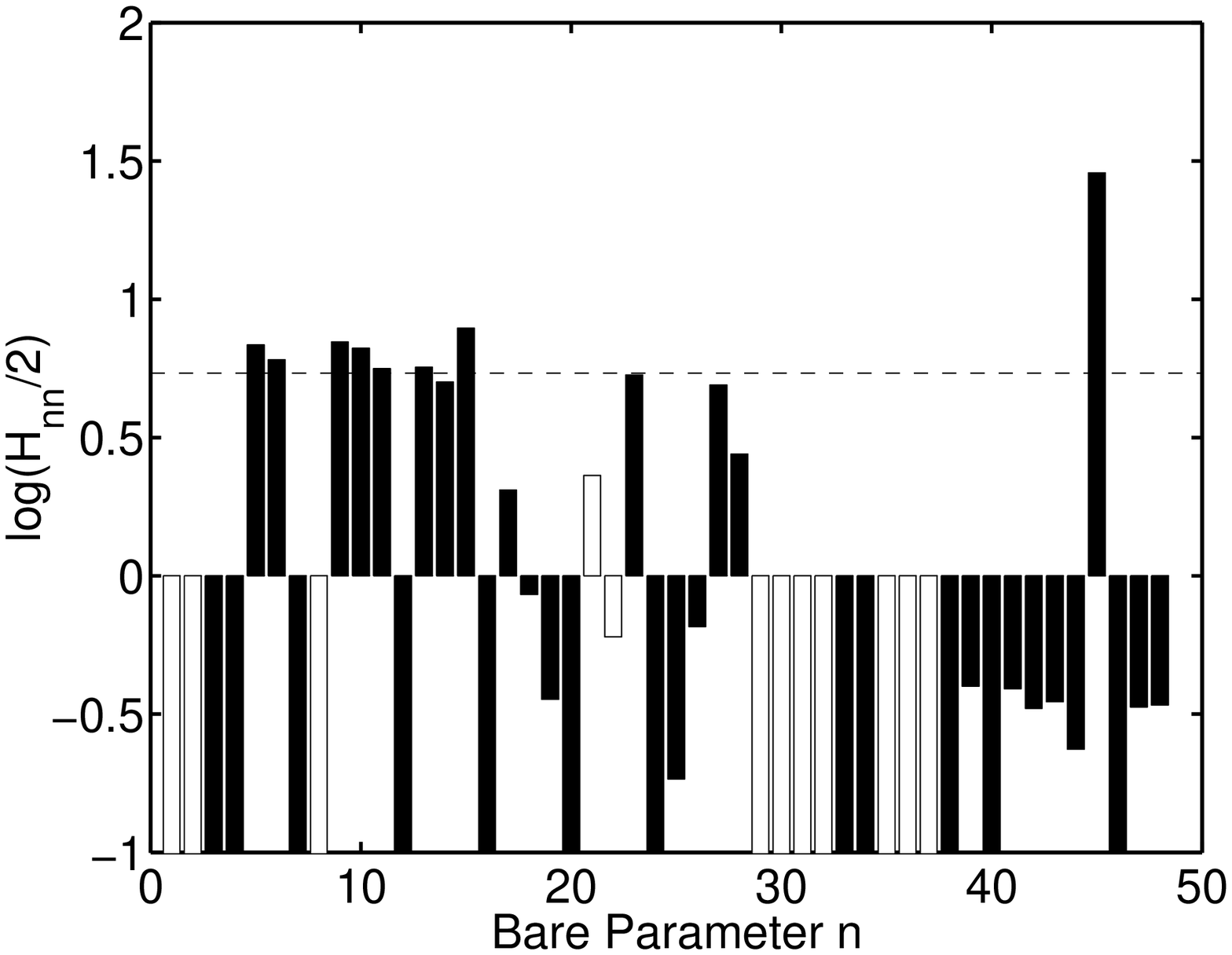}} \\
    \end{tabular}
\end{center}
\caption{Plot showing the logarithm of the robustness (equation \eref{eqn:robust}) for each parameter of the PC12 model.  In both (a) and (b), bars that fall
below the dotted line correspond to robust parameters and bars above the dotted line are nonrobust parameters (see ``Experimental Section'' for more details).
Note the different scales in (a) and (b). Color coding of the networks indicates experimental data sets that were present for Hessian computation; proteins for
which EGF--response data were used are colored in magenta and those for which NGF--response data were used are colored in green. (a)~In the case where we have
constraining data available for many proteins in the network, we are probing the sensitivity of a multifunctional object to parameter variations and find a
very nonrobust system; 37 out of the 48 parameters lie above our line of significance. (b)~When we view the network as an input/output transducer with Erk1/2
activation as the output, we find a very robust result; 39 of the 48 parameters can be varied by a factor of $2$ with no significant effect on the network
output.  White bars in the bar graphs correspond to diagonal elements $H_{nn}$ that have (small) negative values; in these cases the plot displays the log of
the absolute value.} \label{fig:robust}
\end{figure}

\begin{figure}[h]
\begin{center}
\includegraphics[width=8.7cm]{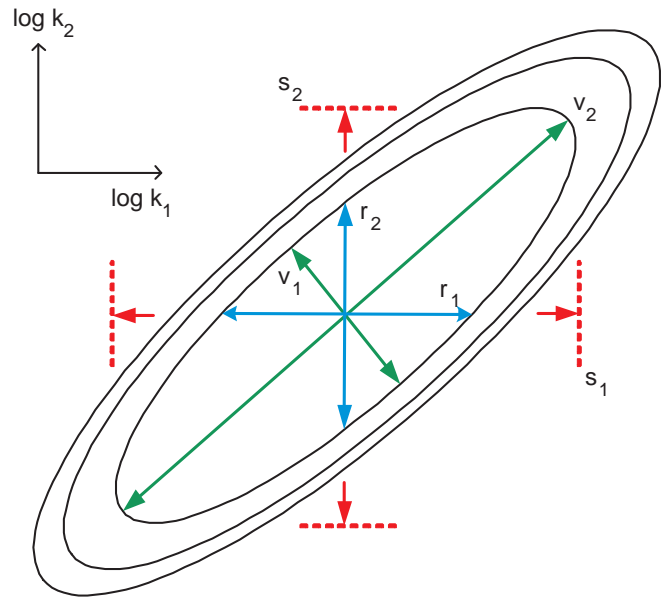}
\end{center}
\caption{Schematic illustrating the difference between robustness range, parameter error bars, and stiff and soft eigendirections, all of which are forms of
parameter variation discussed in the text.  The black ellipses represent lines of equal cost for a simple quadratic cost function with only two parameters,
$k_1$ and $k_2$.  The best fit (minimum cost) is positioned where the green and blue axes cross.  The blue axes (whose tips are labeled $r_1$ and $r_2$) show
the robustness range of the two parameters, for a quantitative robustness definition similar to the one described in the text.  The red dotted lines and arrows
(labeled $s_1$ and $s_2$) give the error bars for the parameters.  Notice that the parameter error bars are larger than the robustness range: unless the rate
constants show no covariance (the axes of the ellipse coincide with the black axes labeled $\log k_1$ and $\log k_2$) this will be the case. The green axes
show the range of the two eigenparameters; notice that one eigenparameter ($v_1$) has a much smaller range of variation than the error bar or robustness range
of either parameter, and one eigenparameter ($v_2$) has a much larger range of variation.  We would call $v_1$ stiff and $v_2$ soft.  This schematic differs
from the actual behavior of our model near the cost minimum in three significant ways. First, the schematic shows only two dimensions, rather than 48 for the
PC12 model. Second, the model shows perfect ellipses, ignoring the nonlinearities that make stochastic sampling a necessity. Third, typical elliptical contours
in the PC12 model are vastly more eccentric (needle-like): the robustness range (blue) is typically less than a factor of two, while the error bars (red) are
all larger than a factor of 1000.} \label{fig:ellipse}
\end{figure}

\begin{figure}[h]
    \begin{center}
        \begin{tabular}{lc}
        {\bf \Large (a)} & \resizebox{8.8cm}{!}
      {\includegraphics[width=8.7cm,angle=-90]{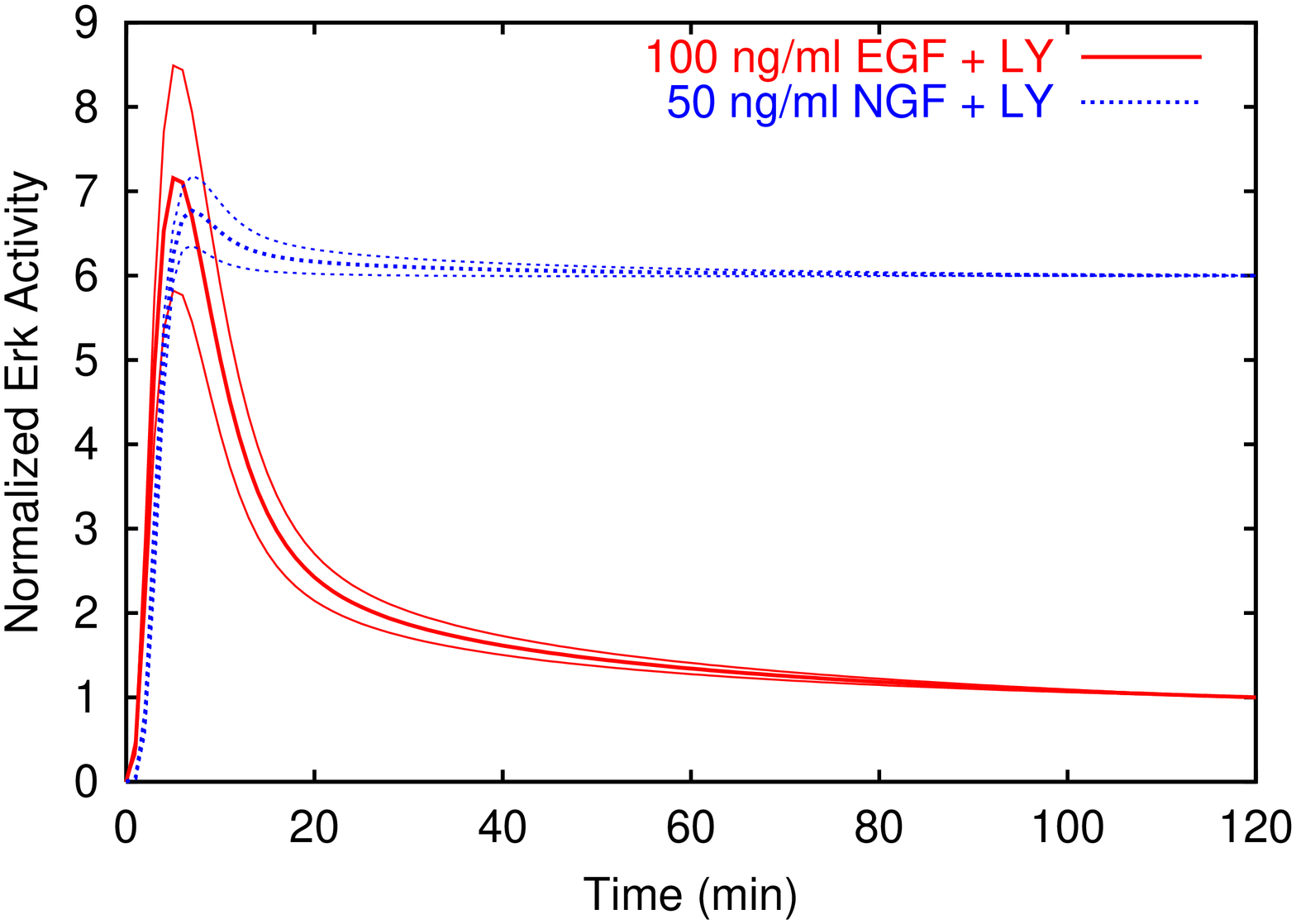}}
      \\
      \vspace{0.25in} \\
      {\bf \Large (b)} & \resizebox{8.8cm}{!}
      {\includegraphics[width=8.7cm,angle=-90]{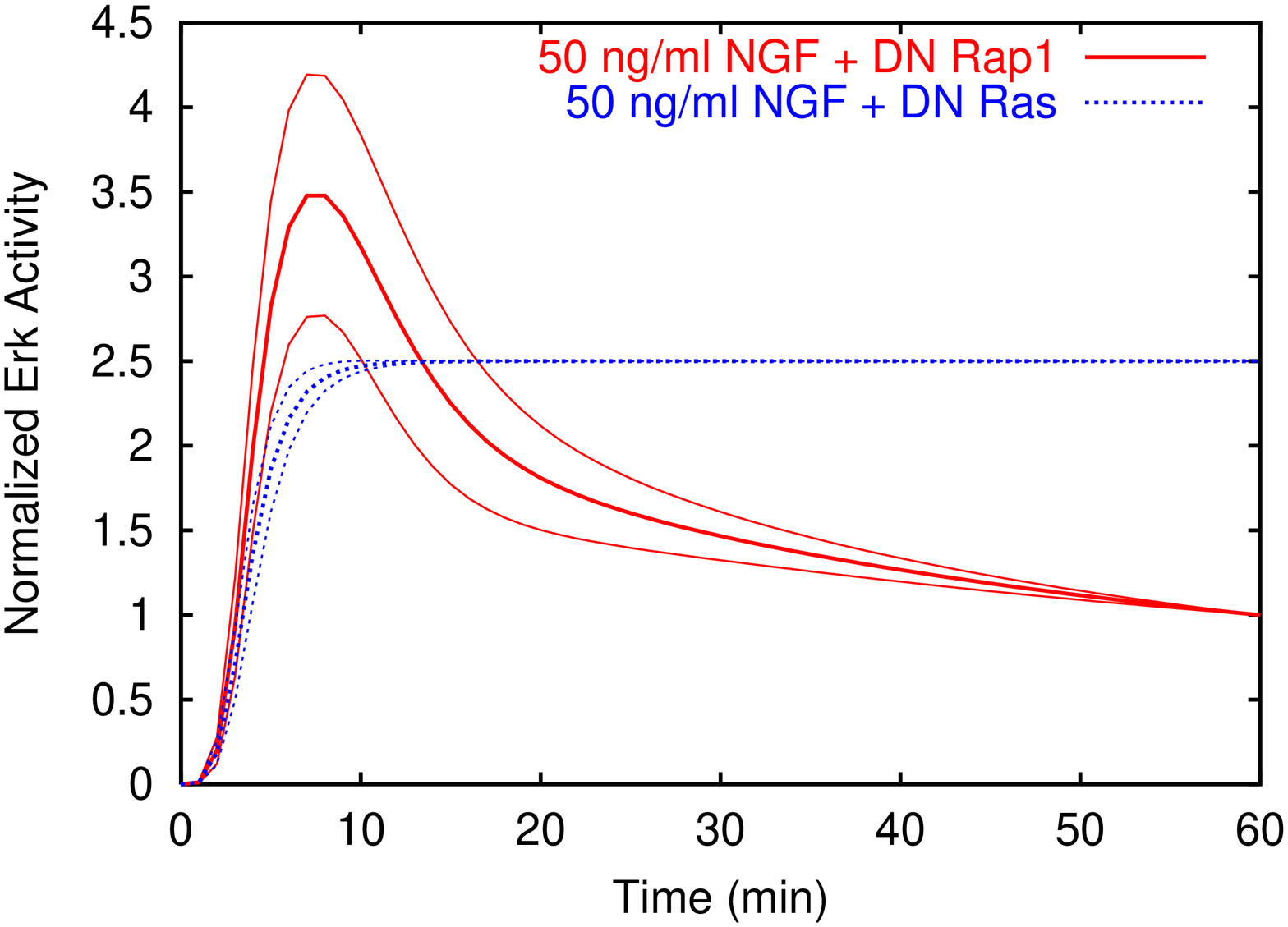}}
      \\
        \end{tabular}
\end{center}
\caption{Predictions of the PC12 cell model. (a)~Prediction of the effects of PI3K inhibition on Erk1/2 activity in response to EGF (red curve) and NGF (blue
curve) stimulation.  The data has been normalized so that all curves pass through the same point at 90 minutes; this is because the units for the biological
fitting data are relative (like fold activity) rather than absolute (like molar).  Regardless, the transient nature of the signal is absolutely clear. These
predicted curves were generated from the same samples that generated Fig.~2a. LY is LY294002, a chemical inhibitor of PI3K. Notice, in sharp contrast to the
ill-determined rate constants, our ensemble makes rather definite predictions for the time series. Notice also that the prediction is counter to our initial
hypothesis, that PI3K was important in downregulating Erk. (b)~Prediction of the effects of dominant--negative (DN) Ras and Rap1 on NGF--mediated Erk1/2
activity.  The data have been normalized as in (a). Notice that at the $1\sigma$ level (shown), Rap1 inhibition causes a transient Erk profile, but at the
$2\sigma$ level the result is somewhat ambiguous.  This observation points to the importance of the ensemble in determining which experimental manipulations
are ``close calls,'' {\it i.e.} model predictions which are not conclusive. No such information is contained in a single set of parameters. Panel (b) should be
compared to data of \citeasnoun{york98}, where the authors show that DN Ras affects early--time NGF--mediated Erk1/2 signaling but not its saturation and DN
Rap1 affects the sustained response of the signal without disrupting early--time activation.} \label{fig:predict}
\end{figure}

\begin{figure}[h]
\begin{center}
\end{center}
\caption{Western blot of PC12 cells treated with 10 $\mu$M LY294002 and either 100 ng/ml EGF or 50 ng/ml NGF. Note, as predicted in figure~4(a) (and in
contradiction to our intuition), the inhibition of PI3K does not switch transient Erk1/2 phosphorylation to sustained activity in response to EGF treatment,
nor does it prevent sustained Erk1/2 phosphorylation in response to NGF.} \label{fig:western}
\end{figure}

\clearpage

\end{document}